\documentclass[aps,prd,superscriptaddress,notitlepage,onecolumn,11pt,a4paper,nofootinbib]{revtex4-1}
\usepackage{amsmath}
\usepackage{bm}
\usepackage{times}
\usepackage{braket}
\usepackage{color,graphicx}
\usepackage[T1]{fontenc}
\usepackage[utf8]{inputenc}
\usepackage{hyperref}
\usepackage[normalem]{ulem}
\usepackage{array}

\definecolor{nicered}{rgb}{0.7,0.1,0.1}
\definecolor{nicegreen}{rgb}{0.1,0.5,0.1}

\newcommand{\bea}{\begin{eqnarray}}
\newcommand{\eea}{\end{eqnarray}}

\hypersetup{colorlinks,citecolor= nicegreen, linkcolor= nicered}

\begin{document}
\title{Leptoquark mechanism of neutrino masses within the grand unification framework}

\author{Ilja Dor\v sner} \email[Electronic address:]{dorsner@fesb.hr}
\affiliation{University of Split, Faculty of Electrical Engineering, Mechanical Engineering and Naval Architecture in Split (FESB), Ru\dj era Bo\v skovi\' ca 32, 21000 Split, Croatia}

\author{Svjetlana Fajfer} \email[Electronic
address:]{svjetlana.fajfer@ijs.si} 
\affiliation{Department of Physics,
  University of Ljubljana, Jadranska 19, 1000 Ljubljana, Slovenia}
\affiliation{J.\ Stefan Institute, Jamova 39, P.\ O.\ Box 3000, 1001
  Ljubljana, Slovenia}

\author{Nejc Ko\v snik} 
\email[Electronic address:]{nejc.kosnik@ijs.si}
\affiliation{J.\ Stefan Institute, Jamova 39, P.\ O.\ Box 3000, 1001
  Ljubljana, Slovenia}
\affiliation{Department of Physics,
  University of Ljubljana, Jadranska 19, 1000 Ljubljana, Slovenia}


\begin{abstract}
We demonstrate viability of the one-loop neutrino mass mechanism within the framework of grand unification when the loop particles comprise scalar leptoquarks (LQs) and quarks of the matching electric charge. This mechanism can be implemented in both supersymmetric and non-supersymmetric models and requires the presence of at least one LQ pair. The appropriate pairs for the neutrino mass generation via the up-type and down-type quark loops are $S_3$--$R_2$ and $S_{1,\,3}$--$\tilde{R}_2$, respectively. We consider two phenomenologically distinct regimes for the LQ masses in our analysis. First regime calls for very heavy LQs in the loop. It can be naturally realised with the $S_{1,\,3}$--$\tilde{R}_2$ scenarios when the LQ masses are roughly between $10^{12}$\,GeV and $5 \times 10^{13}$\,GeV. These lower and upper bounds originate from experimental limits on partial proton decay lifetimes and perturbativity constraints, respectively. Second regime corresponds to the collider accessible LQs in the neutrino mass loop. That option is viable for the $S_3$--$\tilde{R}_2$ scenario in the models of unification that we discuss. If one furthermore assumes the presence of the type II see-saw mechanism there is an additional contribution from the $S_3$--$R_2$ scenario that needs to be taken into account beside the type II see-saw contribution itself. We provide a complete list of renormalizable operators that yield necessary mixing of all aforementioned LQ pairs using the language of $SU(5)$. We furthermore discuss several possible embeddings of this mechanism in $SU(5)$ and $SO(10)$ gauge groups.
\end{abstract}


\maketitle

\section{Introduction}
\label{sec:Introduction}

Leptoquarks (LQs) are colored states that couple quarks to leptons. They can thus yield novel physical processes such as proton decay or help explain experimentally observed phenomena that cannot be successfully addressed within the Standard Model (SM) of elementary particle physics. For example, neutrino masses of Majorana nature can be generated through the one-loop level processes if one introduces at least two particular scalar LQ multiplets~\cite{Chua:1999si,Mahanta:1999xd} to the SM particle content. It is our intention to investigate a viability of this particular mechanism within a context of grand unification. This is where the LQs first emerged after all~\cite{Pati:1973uk,Pati:1974yy,Fritzsch:1974nn,Georgi:1974sy}. For exhaustive lists of references on the LQ phenomenology one can consult reviews on the subject~\cite{Davidson:1993qk,Hewett:1997ce,Nath:2006ut,Dorsner:2016wpm} or turn to the numerous studies of specific aspects of the LQ related physics~\cite{Shanker:1981mj,Shanker:1982nd,Buchmuller:1986iq,Buchmuller:1986zs,Hewett:1987yg,Leurer:1993em,Leurer:1993qx}. The one-loop contributions towards neutrino masses that we study have been considered extensively in the literature~\cite{Chua:1999si,Mahanta:1999xd,AristizabalSierra:2007nf,Helo:2015fba,Pas:2015hca,Hagedorn:2016dze,Cheung:2016fjo}. Our intention, in contrast to the existing studies, is to analyse possibilities to have a more fundamental origin of this mechanism and to provide several realistic examples.

To start, we present an overview of the most salient features of this mechanism. Only then do we proceed to discuss two distinct implementations of this approach to address the issue of neutrino mass within the grand unification framework. We list the transformation properties of scalar LQs under the SM gauge group in Table~\ref{tab:LQs}. We adopt symbolic notation to represent LQ multiplets~\cite{Buchmuller:1986zs}. We also denote a given representation with the associated dimensionality whenever possible. To single out a particular electric charge eigenstate from a given LQ multiplet we use superscripts~\cite{Dorsner:2016wpm}. For example, $S_3$ comprises three electric charge eigenstates that we label $S_3^{4/3}$, $S_3^{1/3}$, and $S_3^{-2/3}$. This fixes the hypercharge normalisation we use throughout the manuscript.  
\begin{table}[tbp]
\centering
\begin{tabular}{|c|c|c|c|}
\hline
LEPTOQUARK & $(SU(3),SU(2),U(1))$ & $SU(5)$ & $SO(10)$\\
\hline 
$S_3$ & $(\overline{\mathbf{3}},\mathbf{3},1/3)$ & $\overline{\mathbf{45}}$ & $\mathbf{120},\,\overline{\mathbf{126}}$ \\
$R_2$ & $(\mathbf{3},\mathbf{2},7/6)$ & $\overline{\mathbf{45}},\, \overline{\mathbf{50}}$ & $\mathbf{120},\,\overline{\mathbf{126}}$\\
$\tilde{R}_2$ & $(\mathbf{3},\mathbf{2},1/6)$ & $\mathbf{10},\,\mathbf{15}$ & $\mathbf{120},\,\overline{\mathbf{126}}$\\
$\tilde{S}_1$ & $(\overline{\mathbf{3}},\mathbf{1},4/3)$ & $ \mathbf{45}$ & $\mathbf{120}$\\
$S_1$ & $(\overline{\mathbf{3}},\mathbf{1},1/3)$ & $\overline{\mathbf{5}},\, \overline{\mathbf{45}},\, \overline{\mathbf{50}}$ & $\mathbf{10},\,\mathbf{120},\,\overline{\mathbf{126}}$\\
\hline
\end{tabular}
\caption{\label{tab:LQs} Transformation properties of scalar LQs under the SM gauge group. The list of the most relevant $SU(5)$ ($SO(10)$) representations that accommodate them is presented in the third (fourth) column. We assume the standard embedding of $U(1)$ charges in $SO(10)$.}
\end{table} 

The mechanism we want to study, in its minimal form, requires the presence of one scalar multiplet that transforms as $\tilde{R}_2$ and another one that has the transformation properties of either $S_1$ or $S_3$ in addition to the SM particle content. The following two features are crucial if one is to generate neutrino mass(es) at the one-loop level. Firstly, $\tilde{R}^{-1/3}_{2}$ ($S_1$ and $S_3^{1/3}$) can couple neutrinos to the right-chiral (left-chiral) down-type quarks. The relevant parts of the Yukawa interactions are  
\begin{align}
\label{eq:main_Y_D}
\mathcal{L}_Y \supset &-\tilde{y}^{RL}_{2}\bar{d}_{R} \tilde{R}_{2}^{a}\epsilon^{ab}L_{L}^{b}-y^{LL}_1 \bar{Q}_{L}^{C\,a} S_{1} \epsilon^{ab}L_{L}^{b}- y^{LL}_{3}\bar{Q}_{L}^{C\,a} \epsilon^{ab} (\tau^k S^k_{3})^{bc} L_{L}^{c}-y_{D}\bar{Q}_{L}^{a} H^{a} d_{R}+\textrm{h.c.},
\end{align}
where $\tilde{y}_2^{RL}$, $y_1^{LL}$, $y_3^{LL}$, and $y_D$ are $3 \times 3$ matrices in flavor space.\footnote{The chiralities of the quark--lepton pair that the LQ couples to are denoted with the superscript labels of $\tilde{y}_2^{RL}$, $y_1^{LL}$, and $y_3^{LL}$.} $H (\equiv (\mathbf{1},\mathbf{2},1/2))$ is the Higgs boson of the SM, $\tau^k$, $k=1,2,3$, are Pauli matrices, and $a,b,c=1,2$ are the $SU(2)$ group space indices. The couplings of $\tilde{R}_{2}^{-1/3}$, $S_1$, and $S^{1/3}_3$ with the left-chiral neutrinos are $\tilde{y}_2^{RL} \bar{d}_{R}\nu_{L} \tilde{R}_{2}^{-1/3}$, $y_1^{LL} \bar{d}_{L}^{C} \nu_{L} S_{1}$, and $y_3^{LL} \bar{d}_{L}^{C} \nu_{L} S^{1/3}_{3}$, respectively.

Secondly, $\tilde{R}_2$ can mix with either $S_{1}$ or $S_{3}$ through the Higgs boson.
In fact, the LQ pairs $S_1$--$\tilde{R}^{-1/3\,*}_{2}$ or $S_3^{1/3}$--$\tilde{R}^{-1/3\,*}_{2}$ should mix in order for the mechanism to work. In the latter case the states $S_3^{-2/3}$ and $\tilde{R}^{2/3\,*}_{2}$ also mix. The relevant parts of the scalar interactions are
\begin{equation}
\label{eq:main_H_D}
\mathcal{L}_\mathrm{scalar} \supset - \lambda_1 \tilde{R}_{2}^{\dagger\,a} H^{a} S^{\dagger}_1 - \lambda_3 \tilde{R}_{2}^{\dagger\,a} (\tau^k S^{\dagger\,k}_{3})^{ab} H^{b} +\textrm{h.c.},
\end{equation}
where $\lambda_1$ and $\lambda_3$ are dimensionful parameters that we take to be real for simplicity. We denote the squared-masses of the two physical LQs of the $1/3$ electric charge with $m^2_{\mathrm{LQ}\,1}$ and $m^2_{\mathrm{LQ}\,2}$ regardless of whether these states originate from the $S_1$--$\tilde{R}^{-1/3\,*}_{2}$ or $S_3^{1/3}$--$\tilde{R}^{-1/3\,*}_{2}$ combination. The angle that diagonalises $2 \times 2$ squared-mass matrix $m^2_1$ ($m^2_3$) for the $S_1$--$\tilde{R}^{-1/3\,*}_{2}$ ($S_3^{1/3}$--$\tilde{R}^{-1/3\,*}_{2}$) pair is labeled $\theta_1$ ($\theta_3$). The squared-mass matrices $m^2_1$ and $m^2_3$ take the form
\begin{equation*}
m^2_{1,\,3}=\left( \begin{array}{cc}
m^2_{11}   &  \lambda_{1,\,3} \langle H \rangle \\
\lambda_{1,\,3} \langle H \rangle & m^2_{22}
\end{array} \right),
\end{equation*}
where $\langle H \rangle$ represents a vacuum expectation value (VEV) of electrically neutral component of the SM Higgs field. Here, $m^2_{11}$ and $m^2_{22}$ are the squares of would-be masses of $S_1$ and $\tilde{R}^{-1/3\,*}_{2}$ or $S_3^{1/3}$ and $\tilde{R}^{-1/3\,*}_{2}$ if there was no mixing whatsoever. The angles $\theta_1$  and $\theta_3$ are defined through
\begin{equation}
\label{eq:theta}
\tan 2 \theta_{1,\,3} = \frac{2 \lambda_{1,\,3} \langle H \rangle}{m^2_{11}-m^2_{22}}.
\end{equation}

\begin{figure}[bp]
\centering
\begin{tabular}{cc}
\includegraphics[scale=1]{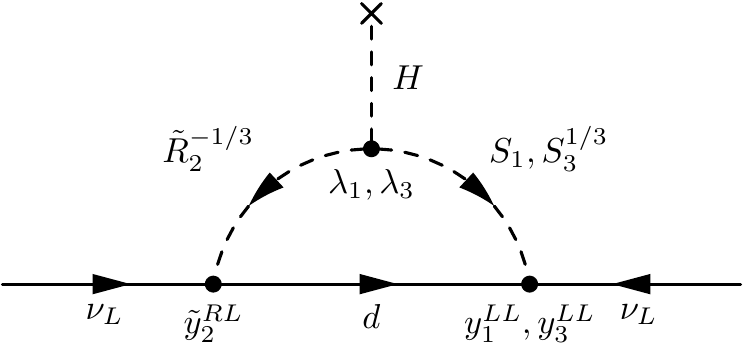} \quad & \quad \includegraphics[scale=1]{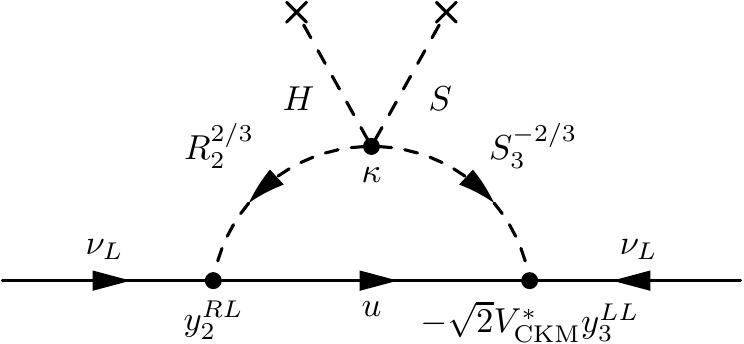}
\end{tabular}
\caption{\label{fig:nu_loop} The one-loop neutrino mass diagrams for the $S_{1,\,3}$--$\tilde{R}_{2}$ and $S_3$--$R_{2}$ scenarios in the left and right panels, respectively. See text for full details.}
\end{figure}

The mechanism is very economical since the same scalar field $H$, upon the electroweak symmetry breaking, provides masses for the SM charged fermions and introduces a mixing term for the LQs. The particles that propagate in the loop that generates neutrino Majorana mass(es) are the down-type quarks and scalar LQs of the matching electric charge. The associated one-loop Feynman diagrams are presented in the left panel of Fig.~\ref{fig:nu_loop}.
The effective neutrino mass matrix in the basis of the physical down-type quarks and LQs reads~\cite{AristizabalSierra:2007nf}
\begin{equation}
\label{eq:m_N}
\begin{split}
(m_N)_{\alpha \beta}&=\frac{3 \sin 2 \theta_{1,\,3}}{32 \pi^2} \sum_{\delta=1,2,3} m_\delta \left[\frac{\log x_{1\delta}}{1-x_{1\delta}} - \frac{\log x_{2\delta}}{1-x_{2\delta}}\right]\left\{(\tilde{y}_2^{RL})_{\delta \alpha} (y_{1,\,3}^{LL})_{\delta \beta}+(\tilde{y}_2^{RL})_{\delta \beta} (y_{1,\,3}^{LL})_{\delta \alpha}\right\}\\
&\approx  \frac{3 \sin 2 \theta_{1,\,3}}{32 \pi^2}  \log\frac{m^2_{\mathrm{LQ}\,2}}{m^2_{\mathrm{LQ}\,1}} \sum_{\delta=1,2,3} m_\delta \left\{(\tilde{y}_2^{RL})_{\delta \alpha} (y_{1,\,3}^{LL})_{\delta \beta}+(\tilde{y}_2^{RL})_{\delta \beta} (y_{1,\,3}^{LL})_{\delta \alpha}\right\},
\end{split}
\end{equation}
where $(m_1,\,m_2,\,m_3)=(m_d,\,m_s,\,m_b)= \Braket{H} ((y_{D})_{11},\,(y_{D})_{22},\,(y_{D})_{33})$ are the down-type quark masses, $\alpha,\beta,\delta=1,2,3$ are flavor indices, and $x_{i\delta} = m_\delta^2/m^2_{\mathrm{LQ}\,i}$.


Before we proceed we have one specific comment with regard to the previous discussion. It concerns a possibility that the fermions that propagate in the neutrino mass loop are the up-type quarks instead of the down-type quarks. This seems to be a viable possibility if one starts with the $R_{2}$--$S_3$ combination. The most essential Yukawa interactions for this scenario are
\begin{align}
\label{eq:main_Y_U}
\mathcal{L}_Y \supset &-y^{RL}_{2}\bar{u}_{R} R_{2}^{a}\epsilon^{ab}L_{L}^{b}- y^{LL}_{3}\bar{Q}_{L}^{C\,a} \epsilon^{ab} (\tau^k S^k_{3})^{bc} L_{L}^{c}-y_{U}\bar{u}_{R} H^{a} \epsilon^{ab} Q_{L}^{b} +\textrm{h.c.},
\end{align}
where $y^{RL}_{2}$ and $y_U$ are $3 \times 3$ matrices in flavor space. The couplings of $R_{2}^{2/3}$ and $S^{-2/3}_3$ with the SM fermions are $y_2^{RL} \bar{u}_{R}\nu_{L} R_{2}^{2/3}$ and $- \sqrt{2} (V^*_\mathrm{CKM} y_3^{LL}) \bar{u}_{L}^{C} \nu_{L} S^{-2/3}_{3}$, where $V_\mathrm{CKM}$ is a Cabibbo-Kobayashi-Maskawa mixing matrix. These couplings, though needed, are not enough to complete the neutrino mass loop since $R_{2}$ and $S_3$ cannot couple directly through $H$ at renormalizable level. One possible remedy is to have an operator of dimension five of the form $R^\dagger_{2} S^\dagger_3 H H H$ that is suppressed by an appropriate scale. Another possibility is to mix $R^{2/3}_{2}$ with $\tilde{R}^{2/3}_{2}$ and $\tilde{R}^{2/3}_{2}$ with $S^{-2/3\,*}_3$ through the SM Higgs fields. This would induce a mixing between $R^{2/3}_{2}$ and $S^{-2/3\,*}_3$ but only if all three multiplets, i.e., $R_{2}$, $\tilde{R}_{2}$, and $S_3$, are present in the set-up~\cite{AristizabalSierra:2007nf}. Third option is to have one additional scalar $S( \equiv (\mathbf{1},\mathbf{3},1))$ that acquires a VEV. The tree-level mixing of $R^{2/3}_{2}$ with $S^{-2/3\,*}_3$ is then possible and the off-diagonal entry of the relevant $2 \times 2$ squared-mass matrix is proportional to a product of the VEVs of neutral fields in $S$ and $H$. The scalar interactions that are needed to implement the second and third option are
\begin{equation}
\label{eq:main_H_U}
\mathcal{L}_\mathrm{scalar} \supset - \lambda_3 \tilde{R}_{2}^{\dagger\,a} (\tau^k S^{\dagger\,k}_{3})^{ab} H^{b}- \lambda_2 R_{2}^{\dagger \,a} H^{a} H^{b} \epsilon^{bc} \tilde{R}_{2}^{c} -\kappa_{1\,(2)} R_{2}^{\dagger \,a} H^{a\,(c)} (\tau^k S^{\dagger\,k}_{3})^{bc} (\tau^l S^{l})^{cb\,(ab)}+\textrm{h.c.},
\end{equation}
where $\lambda_2$ is a dimensionful parameter, whereas $\kappa_{1}$ and $\kappa_{2}$ are both dimensionless parameters. One can trivially  adapt neutrino mass relation in Eq.~\eqref{eq:m_N} to the up-type quark scenario. Let us denote with $\theta_2$ the mixing angle between $R^{2/3}_{2}$ and $S^{-2/3\,*}_3$ states. The
squared-mass matrix $m^2_2$ for the $R^{2/3}_{2}$--$S^{-2/3\,*}_3$ pair takes the form
\begin{equation*}
m^2_2=\left( \begin{array}{cc}
m^2_{11}   &  2 \kappa \langle H \rangle \langle S \rangle \\
2 \kappa \langle H \rangle \langle S \rangle & m^2_{22}
\end{array} \right),
\end{equation*}
where $\langle S \rangle$ represents the VEV of electrically neutral component of $S$ and $\kappa=\kappa_1+\kappa_2$. All one needs to do is to first evaluate $\theta_2$ by replacing $\lambda_{1,\,3} \langle H \rangle$ with $2 \kappa \langle H \rangle \langle S \rangle$ in Eq.~\eqref{eq:theta} and then substitute $\theta_{1,\,3}$, $\tilde{y}_2^{RL}$, and $y_{1,\,3}^{LL}$ with $\theta_2$, $y_2^{RL}$, and $- \sqrt{2} (V^*_\mathrm{CKM} y_3^{LL})$, respectively, in Eq.~\eqref{eq:m_N}. The one-loop Feynman diagram that corresponds to the $S$--$H$ induced mixing of the $R^{2/3}_{2}$--$S^{-2/3\,*}_3$ pair is shown in the right panel of Fig.~\ref{fig:nu_loop}. We will make further comments on this potentially important contribution towards neutrino masses later on.

Our aim is to implement the one-loop neutrino mass mechanism in the framework of grand unification. We accordingly investigate viability of two distinct regimes in Section~\ref{sec:GUT} using mainly the language of $SU(5)$ gauge group. First regime corresponds to a scenario where the LQs behind the neutrino mass generation reside at a very high energy scale. This possibility is discussed in Section~\ref{sec:heavy}. Second regime corresponds to a scenario where the neutrino masses are generated with the Large Hadron Collider (LHC) accessible scalar LQs. We demonstrate viability of that scenario in Section~\ref{sec:light}. The summary of our findings is presented in Section~\ref{sec:Conclusions}.

\section{Grand unification vs.\ one-loop neutrino mass}
\label{sec:GUT}

Let us proceed with a realistic implementation of the one-loop neutrino mass mechanism with scalar LQs in the grand unification framework. We primarily use the language of the $SU(5)$ gauge group in what follows. The SM fermions reside in $\mathbf{10}_\alpha$ and $\overline{\mathbf{5}}_\alpha$ of $SU(5)$, where $\alpha(=1,2,3)$ is a flavor index~\cite{Georgi:1974sy}. The exact decompositions of $\mathbf{10}_\alpha$ and $\overline{\mathbf{5}}_\alpha$ under $SU(3) \times SU(2) \times U(1)$ of the SM read $\mathbf{10}_\alpha \equiv (\mathbf{1},\mathbf{1},1)_\alpha \oplus(\overline{\mathbf{3}},\mathbf{1},-2/3)_\alpha
\oplus(\mathbf{3},\mathbf{2},1/6)_\alpha=(e^C_\alpha,u^C_\alpha,Q_\alpha)$ and $\overline{\mathbf{5}}_\alpha \equiv (\mathbf{1},\mathbf{2},-1/2)_\alpha \oplus
(\overline{\mathbf{3}},\mathbf{1},1/3)_\alpha=(L_\alpha,d^C_\alpha)$, respectively. Possible embeddings of scalar LQs in the $SU(5)$ representations are presented in Table~\ref{tab:LQs}. We clearly need to have either one $10$- or one $15$-dimensional scalar representation in order to introduce one $\tilde{R}_2$ multiplet in any $SU(5)$ model. Relevant contraction that yields $\tilde{y}_2^{RL} \bar{d}_{R}\nu_{L}\tilde{R}_{2}^{-1/3}$ term when $\tilde{R}_{2}$ is part of $10$-dimensional ($15$-dimensional) representation is $y_{\alpha \beta} \overline{\mathbf{5}}_\alpha \overline{\mathbf{5}}_\beta \mathbf{10}$ ($y_{\alpha \beta} \overline{\mathbf{5}}_\alpha \overline{\mathbf{5}}_\beta \mathbf{15}$). We identify $(\tilde{y}_2^{RL})_{\alpha \beta}$ to be $-y_{\alpha \beta}/\sqrt{2}$, where $y_{\alpha \beta}$ are elements of an antisymmetric (symmetric) complex matrix in the case when $\tilde{R}_{2}$ originates from $10$-dimensional ($15$-dimensional) representation.

The mass mechanism that we discuss can also be implemented in the $SO(10)$ framework. See Table~\ref{tab:LQs} for the standard embedding of scalar LQs in the $SO(10)$ representations. In particular, if $\tilde{R}_{2}$ originates from $120$-dimensional ($126$-dimensional) representation of $SO(10)$ the relevant couplings to the SM fermions will be antisymmetric (symmetric) in flavor space. These properties thus closely mirror the $SU(5)$ flavor structure of the $\tilde{R}_{2}$ couplings. The associated $SO(10)$ operators are $y_{\alpha \beta} \mathbf{16}_\alpha \mathbf{16}_\beta \mathbf{120}$ and $y_{\alpha \beta} \mathbf{16}_\alpha \mathbf{16}_\beta \overline{\mathbf{126}}$, where we assume that one $16$-dimensional $SO(10)$ representation comprises one generation of the SM fermions and one right-chiral neutrino.

The origin of the term $y_3^{LL} \bar{d}_{L}^{C} \nu_{L} S^{1/3}_{3}$ is unique in $SU(5)$ as can be seen from Table~\ref{tab:LQs}. Namely, $S_3$ resides in a $45$-dimensional representation and the relevant contraction that generates aforementioned couplings is $y^{45}_{\alpha \beta} \mathbf{10}_\alpha \overline{\mathbf{5}}_\beta \overline{\mathbf{45}}$. One can thus identify $y_3^{LL}$ with $y^{45}/\sqrt{2}$, where $y^{45}$ is related to the masses of the charged fermions and down-type quarks as we show in the next paragraph. The situation with $R_2$ seems more involved since $R_2$ can belong to either $45$- or $50$-dimensional representation. But, if it originates from $50$-dimensional representation it cannot couple to the left-chiral neutrinos. This then leaves $45$-dimensional representation as the only possible source of $R_2$. The operator $y_2^{RL} \bar{u}_{R}\nu_{L} R_{2}^{2/3}$ thus originates from $y^{45}_{\alpha \beta} \mathbf{10}_\alpha \overline{\mathbf{5}}_\beta \overline{\mathbf{45}}$, where $y_2^{RL}$ can be identified with $-y^{45}$. The flavor structure of relevant interactions of $S_3$ and $R_2$ with the SM fermions in $SO(10)$ depends on whether these states originate from $120$- or $126$-dimensional representation. In the former (latter) case the relevant couplings to the SM fermions are antisymmetric (symmetric) in the flavor basis.

To generate viable charged fermion masses the minimal $SU(5)$ scenario needs to include one $5$-dimensional scalar representation beside the $45$-dimensional one~\cite{Georgi:1979df}. We denote VEVs of $\mathbf{5} \equiv \mathbf{5}^i$ and $\mathbf{45} \equiv \mathbf{45}^{ij}_k$ with $\langle\mathbf{5}^5\rangle= v_5 / \sqrt{2}$ and $\langle\mathbf{45}^{15}_{1}\rangle= \langle\mathbf{45}^{2 5}_{2}\rangle=\langle\mathbf{45}^{3 5}_{3}\rangle = v_{45}/\sqrt{2}$, where $i,j,k=1,\ldots,5$ are the $SU(5)$ indices. The minimal set of contractions that generates mass matrices of the SM charged fermions comprises three operators: $y^{45}_{\alpha \beta} \mathbf{10}_\alpha \overline{\mathbf{5}}_\beta \overline{\mathbf{45}}$, $y^{5}_{\alpha \beta} \mathbf{10}_\alpha \overline{\mathbf{5}}_\beta \overline{\mathbf{5}}$, and $\bar{y}_{\alpha \beta} \mathbf{10}_\alpha \mathbf{10}_\beta \mathbf{5}$. The $3 \times 3$ mass matrices for the down-type quarks, charged leptons, and the up-type quarks are
\begin{align}
\label{eq:m_D}
m_D&=- y^{45} v_{45}- y^{5} v_{5}/2,\\
\label{eq:m_E}
m^T_E&=3 y^{45} v_{45}- y^{5} v_{5}/2,\\
\label{eq:m_U}
m_U&=\sqrt{2}(\bar{y}+\bar{y}^{T}) v_{5},
\end{align}
where all the VEVs are taken to be real. The VEV normalization yields $v_5^2/2+12 v_{45}^2=v^2$, where $v(=246\,\mathrm{GeV})$ is the electroweak VEV~\cite{Dorsner:2011ai}. The $SU(5)$ symmetry thus dictates that $y^{45} \equiv \sqrt{2} y_3^{LL} =-y_2^{RL}=(m^T_E-m_D)/(4 v_{45})$. The term $y_1^{LL} \bar{d}_{L}^{C} \nu_{L} S_{1}$ originates from $y^{5}_{\alpha \beta} \mathbf{10}_\alpha \overline{\mathbf{5}}_\beta \overline{\mathbf{5}}$ and $y^{45}_{\alpha \beta} \mathbf{10}_\alpha \overline{\mathbf{5}}_\beta \overline{\mathbf{45}}$ for $S_{1} \in \overline{\mathbf{5}}$ and $S_{1} \in \overline{\mathbf{45}}$, respectively. In the former (latter) case one can identify $y_1^{LL}$ with $-y^{5}/\sqrt{2}$ ($y^{45}/2$).

Finally, one needs to provide the mixing term for at least one of the relevant LQ pairs in order to complete the neutrino mass loop. There are two very different regimes for the scalar LQ masses that we can envisage with this in mind. First option is that the LQs behind the neutrino mass generation reside at a very high energy scale. This could provide compliance of the set-up with the experimental bounds on proton decay. The main issue with this regime could be associated with the size of the relevant lepton-quark-LQ couplings. Namely, these couplings might need to be unrealistically large in order to (re)produce neutrino mass scales that are compatible with experimental observations. It turns out that this is not the case and we accordingly demonstrate in Section~\ref{sec:heavy} why and how this particular scenario can be realised within the grand unification frameworks.

Second option is that the scalar LQs are very light. That scenario is especially appealing since the LHC accessible LQs could also affect flavor physics observables. The main difficulty with this particular set-up is to explain observed levels of matter stability.\footnote{For the latest experimental bounds on proton lifetime see, for example, Ref.~\cite{Miura:2016krn}.} Namely, $S_1$ and $S_3$ can both have ``diquark'' couplings that, in combination with the lepton--quark--LQ couplings that are needed to generate neutrino masses, destabilise protons and bound neutrons.\footnote{$R_2$ and $\tilde{R}_2$ are the only scalar LQs of a ``genuine'' kind as they do not possess ``diquark'' couplings.} To avoid conflict with stringent limits on proton lifetime one would need to either forbid or substantially suppress these ``diquark'' operators. This might be very difficult from the model building point of view since unification of matter multiplets dictates common origin of both types of couplings. One would also need to prevent mixing between these LQs and any other LQ in the theory that has ``diquark'' couplings to insure stability of matter. This might also represent a challenge since one needs to mix specific LQ multiplets in order to generate neutrino masses in the first place. We show that both of these issues can be successfully addressed for the $S_3$--$\tilde{R}_2$ and $S_3$--$R_2$ scenarios in Section~\ref{sec:light}. The $S_1$--$\tilde{R}_{2}$ option, on the other hand, is problematic due to difficulty with suppression of the $S_1$ ``diquark'' couplings in the simplest of models and we opt not to discuss it in the light LQ regime. 

\subsection{Heavy leptoquark regime}
\label{sec:heavy}

Let us turn our attention to a scenario where the LQs are heavy. We assume in what follows that all the LQ masses need to be at or exceed $10^{12}$\,GeV to insure proton stability. This is a very conservative estimate since it is certainly above a lower bound that can be extracted from the latest data on proton stability within the $SU(5)$ framework~\cite{Dorsner:2012uz}. We show that the one-loop neutrino masses can be realised in this part of phenomenologically available parameter space if the fermions in the neutrino mass loop are exclusively the down-type quarks.

The mixing angle between either $S_1$ and $\tilde{R}^{-1/3\,*}_{2}$ or $S_3^{1/3}$ and $\tilde{R}^{-1/3\,*}_{2}$ will be rather small if the LQs are heavy. The $S_3^{1/3}$--$\tilde{R}^{-1/3\,*}_{2}$ mixing, in particular, originates in $SU(5)$ from three operators if $\tilde{R}_2$ originates from $15$-dimensional representation. These operators are $\mathbf{45}^{ij}_k \overline{\mathbf{15}}_{jl} \mathbf{45}^{lk}_i$, $\mathbf{45}^{ij}_k \overline{\mathbf{15}}_{jl} \mathbf{45}^{lk}_m \mathbf{24}_{i}^m$, and $\mathbf{5}^i \overline{\mathbf{15}}_{lj} \mathbf{45}^{jk}_i \mathbf{24}^{l}_k$, where $24$-dimensional representation breaks $SU(5)$ down to $SU(3) \times SU(2) \times U(1)$ through a very large VEV of the order of $10^{16}$\,GeV. We list all possible $SU(5)$ operators that generate mixing between the $1/3$ electric charge scalar LQs that are relevant for the loop generated neutrino masses in Table~\ref{table:SU(5)}. For example, the operator $\mathbf{5}^i \overline{\mathbf{15}}_{lj} \mathbf{45}^{jk}_i \mathbf{24}^{l}_k$ produces a mixing coefficient for the $S_3^{1/3}$--$\tilde{R}^{-1/3\,*}_{2}$ pair that is equal to $-5 v_5 v_{24}/(2 \sqrt{2})$, where the VEV of $(\mathbf{1},\mathbf{1},0)$ in $\mathbf{24} \equiv \mathbf{24}^{i}_j$ is $\langle
(\mathbf{1},\mathbf{1},0) \rangle= v_{24}\, \textrm{diag}(2,2,2,-3,-3)$. The angle $\theta_3$ of Eq.~\eqref{eq:theta} can thus be approximated to be at most $\theta_3 \sim (v_5 v_{24})/m_\mathrm{LQ}^2 \approx10^{18}/10^{24} = 10^{-6}$, where $ v_5 \sim \langle H \rangle \approx 10^{2}$\,GeV, $v_{24} \sim \lambda_3 \approx 10^{16}$\,GeV, and $m^2_{11}-m^2_{22} \sim m_\mathrm{LQ}^2 \approx 10^{24}$\,GeV$^2$. The necessary mixing between $S_1(\in \mathbf{5})$ and $\tilde{R}_{2}(\in\mathbf{15})$ can be generated through the contractions of the form $\overline{\mathbf{5}}_i \overline{\mathbf{5}}_j \mathbf{15}^{ij}$ and $\overline{\mathbf{5}}_i \overline{\mathbf{5}}_j \mathbf{15}^{jk} \mathbf{24}_{k}^i$. These, again, yield an angle $\theta_1$ that is comparable in strength to our estimate for $\theta_3$. We can furthermore safely assume that $m_b( \approx 1$\,GeV) contribution dominates the sum in Eq.~\eqref{eq:m_N}. Putting all this together implies that
\begin{equation}
\label{eq:m_N_approx}
m_N \sim \frac{3 \theta_{1,\,3}}{32 \pi^2}  m_b \ln \frac{m^2_{\mathrm{LQ}\,2}}{m^2_{\mathrm{LQ}\,1}} (\tilde{y}_2^{RL} y_{1,\,3}^{LL}) \approx \frac{10^{-6}}{10^{2}}  10^{9}\,\mathrm{eV} (\tilde{y}_2^{RL} y_{1,\,3}^{LL}),
\end{equation} 
where we suppress flavor indices and assume that the mass splitting between LQs is not substantial, i.e., we take that $\ln (m^2_{\mathrm{LQ}\,2}/m^2_{\mathrm{LQ}\,1}) \approx 1$. The approximation of Eq.~\eqref{eq:m_N_approx} shows that the entries in the product $(\tilde{y}_2^{RL} y_{1,\,3}^{LL})$ do not have to be very large to correctly describe the neutrino mass scale. For example, in the non-degenerate normal hierarchy case for the neutrino masses the largest entry on the left side of Eq.~\eqref{eq:m_N_approx} needs to be at the level of $5 \times 10^{-2}$\,eV which would imply that $(\tilde{y}_2^{RL} y_{1,\,3}^{LL}) \sim 5 \times 10^{-3}$.\footnote{For a recent  analysis of neutrino oscillation data see, for example, Ref.~\cite{Capozzi:2016rtj}.} The back-of-the-envelope estimate we present clearly demonstrates viability of this option. Note that there is an upper bound on the heavier of the two LQs in this set-up if one demands perturbativity of Yukawa coupling entries in $\tilde{y}_2^{RL}$ and $y_{1,\,3}^{LL}$ matrices. We find it to be roughly at $5 \times 10^{13}$\,GeV. This implies that the two LQs must reside in relatively narrow mass window from $10^{12}$\,GeV to $5 \times 10^{13}$\,GeV in order to accommodate all the relevant constraints. One can then infer that $\ln (m^2_{\mathrm{LQ}\,2}/m^2_{\mathrm{LQ}\,1}) \lesssim 5$ which is in agreement with our initial assumption.
\begin{table}[htp]
\begin{center}
\begin{tabular}{| c | c | c | c | c |}\hline
\multicolumn{2}{|c|}{}   &  \multicolumn{2}{c|}{$S_1$} & $S_3$\\\cline{3-5}
\multicolumn{2}{|c|}{\raisebox{2.4ex}[0pt]{$SU(5)$}} & $\mathbf{5}$ & $\mathbf{45}$ & $\mathbf{45}$\\\hline
&&$\mathbf{5}^i \overline{\mathbf{10}}_{jk} \mathbf{45}^{jk}_i$&&\\
&&$\overline{\mathbf{5}}_i \overline{\mathbf{5}}_j \mathbf{10}^{jk} \mathbf{24}_{k}^i$&\raisebox{2.4ex}[0pt]{$\mathbf{5}^i \overline{\mathbf{10}}_{jk} \mathbf{45}^{jk}_i$}&$\mathbf{5}^i \overline{\mathbf{10}}_{jk} \mathbf{45}^{jk}_i$\\
&$\mathbf{10}$&$\mathbf{5}^i \overline{\mathbf{10}}_{lj} \mathbf{45}^{jk}_i \mathbf{24}^{l}_k$&\raisebox{2.4ex}[0pt]{$\mathbf{5}^i \overline{\mathbf{10}}_{lj} \mathbf{45}^{jk}_i \mathbf{24}^{l}_k$}&$\mathbf{5}^i \overline{\mathbf{10}}_{lj} \mathbf{45}^{jk}_i \mathbf{24}^{l}_k$\\
&&$\mathbf{5}^i \overline{\mathbf{10}}_{ij} \mathbf{45}^{jk}_l \mathbf{24}^{l}_k$&\raisebox{2.4ex}[0pt]{$\mathbf{5}^i \overline{\mathbf{10}}_{ij} \mathbf{45}^{jk}_l \mathbf{24}^{l}_k$}&$\mathbf{5}^i \overline{\mathbf{10}}_{lm} \mathbf{45}^{lm}_j \mathbf{24}^{j}_i$\\
$\tilde{R}_2$&&$\mathbf{5}^i \overline{\mathbf{10}}_{lm} \mathbf{45}^{lm}_j \mathbf{24}^{j}_i$&\raisebox{2.4ex}[0pt]{$\mathbf{5}^i \overline{\mathbf{10}}_{lm} \mathbf{45}^{lm}_j \mathbf{24}^{j}_i$}&\\\cline{2-5}
&&&$\mathbf{45}^{ij}_k \overline{\mathbf{15}}_{jl} \mathbf{45}^{lk}_i$&\\
&&$\overline{\mathbf{5}}_i \overline{\mathbf{5}}_j \mathbf{15}^{ij}$&$\mathbf{5}^i \overline{\mathbf{15}}_{lj} \mathbf{45}^{jk}_i \mathbf{24}^{l}_k$&\raisebox{2.4ex}[0pt]{$\mathbf{45}^{ij}_k \overline{\mathbf{15}}_{jl} \mathbf{45}^{lk}_i$}\\
&\raisebox{2.4ex}[0pt]{$\mathbf{15}$}&$\overline{\mathbf{5}}_i \overline{\mathbf{5}}_j \mathbf{15}^{jk} \mathbf{24}_{k}^i$&$\mathbf{5}^i \overline{\mathbf{15}}_{ij} \mathbf{45}^{jk}_l \mathbf{24}^{l}_k$&\raisebox{2.4ex}[0pt]{$\mathbf{5}^i \overline{\mathbf{15}}_{lj} \mathbf{45}^{jk}_i \mathbf{24}^{l}_k$}\\
&&&$\mathbf{45}^{ij}_k \overline{\mathbf{15}}_{jl} \mathbf{45}^{lk}_m \mathbf{24}_{i}^m$&\raisebox{2.4ex}[0pt]{$\mathbf{45}^{ij}_k \overline{\mathbf{15}}_{jl} \mathbf{45}^{lk}_m \mathbf{24}_{i}^m$}\\
\hline\end{tabular}
\end{center}
\caption{$SU(5)$ operators that generate mixing between the $1/3$ electric charge scalar LQs if one assumes that the only VEVs in the theory are the ones proportional to $v_{24}$, $v_{45}$, and $v_5$.}
\label{table:SU(5)}
\end{table}

This particular possibility to generate neutrino masses, in our view, has been overlooked in the literature on grand unification. For example, there are two non-supersymmetric models that already have all the necessary ingredients to incorporate this particular scenario. The first model~\cite{Perez:2016qbo} introduces one $10$-dimensional scalar representation on top of $\mathbf{5}$, $\mathbf{24}$, and $\mathbf{45}$ in order to generate neutrino masses through the Zee mechanism~\cite{Zee:1980ai}. The second model~\cite{Dorsner:2007fy} resorts to one $15$-dimensional scalar representation in addition to $\mathbf{5}$, $\mathbf{24}$, and $\mathbf{45}$ in order to generate neutrino masses through the type II see-saw mechanism~\cite{Lazarides:1980nt,Mohapatra:1980yp}. Again, both of these models can accommodate the one-loop mechanism we discuss. 

The heavy LQ regime is also tailor-made for the $SO(10)$ framework. This could especially be beneficial in the scenarios that fail to accommodate neutrino masses in satisfactory manner. Clearly, it is sufficient to have either $120$- or $126$-dimensional representation to introduce LQs that transform as $S_1$, $S_3$, and $\tilde{R}_{2}$. This means that the relevant LQ couplings to the SM matter are always in place if one assumes standard embedding of the SM fermions in $SO(10)$. The only remaining element, i.e., the LQ mixing, depends on the exact scalar sector of the $SO(10)$ theory. We opt to show only one example due to existence of several distinct ways one can realistically break $SO(10)$ down to $SU(3) \times SU(2) \times U(1)$. For example, if we introduce one $210$-dimensional representation to break $SO(10)$ there is an operator of the form $\mathbf{210} \,\mathbf{10} \,\overline{\mathbf{126}}$ that exists regardless whether the theory is supersymmetric or not that yields a mixing between $S_1 (\in \mathbf{10})$ and $\tilde{R}_2 (\in \overline{\mathbf{126}})$, Here, $\mathbf{10}$ and $\overline{\mathbf{126}}$ are scalar representation that generate masses of the SM charged fermions.

\subsection{Light leptoquark regime}
\label{sec:light}

To demonstrate that the collider accessible LQ scenario is a viable option to generate neutrino masses one needs to address the issue of the LQ mixing. Namely, if the genuine LQ states mix with the states that have ``diquark'' couplings it is hard to imagine that matter stability holds at the experimentally observed levels. 
We focus exclusively on a scenario when $\tilde{R}_2$ originates from $15$-dimensional representation. The analysis for the $10$-dimensional representation case is completely analogous as we show in Appendix~\ref{sec:appendix}. The $SU(5)$ scenario under consideration comprises the following scalar representations: $\mathbf{5}$, $\mathbf{15}$, $\mathbf{24}$, and $\mathbf{45}$. We note that $R_2$, $\tilde{R}_2$, and $S_3$ do not have ``diquark'' couplings~\cite{Dorsner:2012nq} at renormalizable level if the charged fermion mass relations are given with Eqs.~\eqref{eq:m_D}, \eqref{eq:m_E}, and \eqref{eq:m_U}. The scalar LQs in this set-up are $S_1^{*} \in\mathbf{5}$, $(\tilde{R}_2^{2/3},\,\tilde{R}_2^{-1/3}) \in \mathbf{15}$, and $(S_3^{4/3\,*},\,S_3^{1/3\,*},\,S_3^{-2/3\,*},\,R^{5/3\,*}_{2},\,R^{2/3\,*}_{2},\,\tilde{S}_1,\,S_1^{*}) \in \mathbf{45}$. All in all, there is one LQ with the $5/3$ charge, two LQs with the $4/3$ charge, three LQs with the $2/3$ charge, and four LQs with the $1/3$ charge.

There are ten non-trivial operators that mix the LQ states of the same electric charge if the only VEVs present are the ones proportional to $v_{24}$, $v_{45}$, and $v_5$. Nine (four) of these contractions affect the $1/3$ ($2/3$) electric charge states. There are no contractions that mix LQs of the $4/3$ electric charge through  these VEVs. The complete list of relevant $SU(5)$ contractions is relegated to Appendix~\ref{sec:appendix}. It turns out that one can write a $4 \times 4$ squared-mass matrix for the $1/3$ electric charge LQs in a block diagonal form where the relevant two blocks are of dimension $2 \times 2$ each. The basis for this matrix is $(S_1^{*}(\mathbf{45}), S_1^{*}(\mathbf{5}), S_3^{1/3\,*}, \tilde{R}_2^{-1/3})$, where we explicitly denote the origin of LQ multiplets that transform as $S_1$ under the SM gauge group. The mixing term between $S_3^{1/3\,*}$ and $\tilde{R}_2^{-1/3}$ we referred to previously as $\lambda_3 \langle H \rangle$ is proportional to a product of $v_{24}$ with $v_5$. Since the LQs of the $4/3$ electric charge do not mix the associated $2 \times 2$ squared-mass matrix has only diagonal entries. These findings guarantee the matter stability even in the presence of the mixing that is needed to generate neutrino masses. Components of $S_3$ and $\tilde{R}_2$ can thus be very light and the resulting neutrino mass matrix is correctly described through the expression of Eq.~\eqref{eq:m_N} due to a block diagonal form of the relevant LQ squared-mass matrix. We briefly postpone the discussion of the mixing between the LQ states with electric charge of $2/3$ since these originate from $R_2$, $\tilde{R}_2$, and $S_3$ multiplets that have no ``diquark'' couplings in this set-up and consequently do not directly affect matter stability.

Let us summarise the main features of the light LQ set-up. $\tilde{R}_{2}$ ($S_3$) originates from $\mathbf{15}$ ($\mathbf{45}$) of $SU(5)$. Again, $\tilde{R}_{2}$ could instead originate from $10$-dimensional representation. The $SU(5)$ symmetry is broken by the VEV of $\mathbf{24}$ down to $SU(3) \times SU(2) \times U(1)$. The Higgs field VEVs that complete the electroweak symmetry breaking reside in both $\mathbf{5}$ and $\mathbf{45}$. The light LQ states are components of $\tilde{R}_{2}$ and $S_3$ and they help generate neutrino masses. Three out of six LQs of the model --- $S_1(\mathbf{45})$, $S_1(\mathbf{5})$, and $\tilde{S}_1$ --- mediate proton decay and need to be heavy. $R_{2}$ can in principle be of an arbitrary mass. Finally, the mass matrix for the up-type quarks is symmetric in accordance with Eq.~\eqref{eq:m_U} which has implications for the gauge-mediated proton decay~\cite{FileviezPerez:2004hn}. We plan to pursue the phenomenology of this set-up in the future works. In this respect, the state $S_3$ with mass close to the LHC reach has been proven to play a beneficial role in addressing hints of lepton flavor universality violation in $b \to s \ell \ell$ and $b \to c \ell \nu$ processes~\cite{Hiller:2014yaa,Barbieri:2015yvd}.

We have, in our analysis, neglected possible VEVs of electrically neutral fields in $15$- and $24$-dimensional representations. The former (latter) field resides in the $(\mathbf{1},\mathbf{3},1)$ ($(\mathbf{1},\mathbf{3},0)$) component of $\mathbf{15}$ ($\mathbf{24}$). We normalize these additional VEVs of $\mathbf{15} \equiv \mathbf{15}^{ij}$ and $\mathbf{24} \equiv \mathbf{24}^i_j$ to be $\langle\mathbf{15}^{55} \rangle= v_{15}$ and $\langle\mathbf{24}^4_4 \rangle = - \langle\mathbf{24}^5_5 \rangle = v_{S}$, respectively. The presence of these VEVs introduces seven additional $SU(5)$ operators that one needs to include in the analysis of the LQ mixing. We list these operators in Appendix~\ref{sec:appendix}.

The one-loop mechanism we discuss is not the only possible contribution towards neutrino masses in the light LQ regime. Note that the VEV of the $15$-dimensional representation can generate neutrino mass(es) of Majorana nature through the type II see-saw mechanism~\cite{Lazarides:1980nt,Mohapatra:1980yp}.\footnote{For explicit realisation of this possibility within a non-supersymmetric $SU(5)$ framework see, for example, Refs.~\cite{Dorsner:2005fq,Dorsner:2005ii}.} More importantly, the up-type quarks can also contribute towards neutrino mass generation since the scalars $R^{2/3}_{2}$, $\tilde{R}_2^{2/3}$, and $S_3^{-2/3\,*}$ mix with or without the VEV of the $15$-dimensional representation~\cite{AristizabalSierra:2007nf}. In the latter case we find that the up-type quark contribution is completely negligible. In the former case the mixing angle $\theta_2$ between $R^{2/3}_{2}$ and $S_3^{-2/3\,*}$ can be sufficiently large even though it cannot possibly exceed $10^{-3}(\sim (v_{15} v_{5})/m^2_\mathrm{LQ})$ if one is to satisfy existing constrains on the size of $v_{15}$ and the direct limits on LQ masses from the LHC searches. We find that the relevant off-diagonal entries $12$, $13$, and $23$ for the symmetric squared-mass matrix of the $2/3$ electric charge LQs in the basis $(S_3^{-2/3\,*}, R^{2/3}_{2}, \tilde{R}_2^{2/3})$ are proportional to $v_{15} v_{5}$, $v_{24} v_{5}$, and $v_{45} v_{5}$, respectively. For the latest direct bounds on LQ masses from the LHC data see, for example, Refs.~\cite{Aad:2015caa,Khachatryan:2015vaa}. Note that $v_{15}$ is bounded from above due to the existing electroweak precision measurements of the so-called $\rho$ parameter~\cite{Olive:2016xmw}. This bound can be avoided if one judiciously adjusts $v_{15}$ and $v_{S}$ to be approximately equal~\cite{Dorsner:2007fy}. This can increase the maximum allowed value of $v_{15}$ but only by a factor of ten. The leading neutrino mass contributions due to propagation of the up-type quarks and the down-type quarks are thus proportional to $\mathcal{O}(10^{-3}) m_t$ and $\mathcal{O}(1) m_b$, respectively, and can be comparable in strength in some parts of the available parameter space. A self-consistent study of neutrino mass(es) should take into account all these contributions if $\tilde{R}_{2}$ originates from $15$-dimensional representation and the VEV proportional to $v_{15}$ is turned on. 
If $\tilde{R}_{2}$ originates from $10$-dimensional representation the only relevant contribution in this regime is due to the down-type quark loop.

\section{Conclusions}
\label{sec:Conclusions}
The one-loop neutrino mass mechanism with scalar LQs in the loop can be embedded within the framework of grand unification regardless of whether the scenario is supersymmetric or not. There exist two distinct regimes for the LQ masses. 

One option is to have heavy LQs in the loops that generate neutrino masses. This option can be naturally realised with the $S_{1,\,3}$--$\tilde{R}_2$ combinations of LQs. The type II see-saw mechanism contribution could also be present and important in some parts of the accessible parameter space. The nice feature of the heavy LQ limit is that the masses of the LQs in the loop can only be between $10^{12}$\,GeV and $5 \times 10^{13}$\,GeV in order to simultaneously avoid experimental limits on partial proton decay lifetimes and still satisfy perturbativity constraints on the lepton-quark-LQ couplings.

The other option is to have collider accessible LQs in the loop. That particular limit can be realised via the loops that contain the down-type quarks and scalars of the matching electric charge that are the mixture of $S_3$ and $\tilde{R}_2$ multiplets. The $S_1$--$\tilde{R}_2$ combination is not a viable option in this limit due to existence of ``diquark'' couplings of $S_1$ in the minimal set-up. If the theory also contains an $SU(2)$ triplet scalar $(\mathbf{1},\mathbf{3},1)$ that gets the VEV one needs to take into account two additional neutrino mass contributions. One is the type II see-saw contribution and the other one is the one-loop contribution due to propagation of the up-type quarks and the scalar states of the same electric charge that originate from the mixture of $S_3$ and $R_2$ multiplets. These three mechanisms can coexist and be of equal importance in some parts of available parameter space. 

We discuss possible origins of scalar LQs that are needed to complete the neutrino mass generating loops using the language of $SU(5)$. We also provide a list of all $SU(5)$ contractions that generate the LQ mixing terms. We furthermore argue that all the necessary ingredients to implement the one-loop neutrino mass mechanism are present in any $SO(10)$ theory with the standard embedding of the matter fields that generates charged fermion masses through renormalizable contractions.


\acknowledgments
I.D.\ would like to thank Jure Zupan for illuminating discussion. This work has been supported in part by Croatian Science Foundation under the project 7118. S.F.\ and N.K.\ acknowledge support of the Slovenian Research Agency (research core funding No.\ P1-0035).


\appendix
\section{$SU(5)$ contractions}
\label{sec:appendix}
The following nine contractions in the $SU(5)$ group space yield mixing terms for the $1/3$ electric charge LQs when the model comprises $5$-, $15$-, $24$-, and $45$-dimensional scalar representations: $\mathbf{5}^i \overline{\mathbf{15}}_{ij} \mathbf{5}^j$, $\mathbf{5}^i \overline{\mathbf{45}}^k_{ij} \mathbf{24}^j_k$, $\mathbf{45}^{ij}_k \overline{\mathbf{15}}_{jl} \mathbf{45}^{lk}_i$, $\mathbf{5}^l \overline{\mathbf{5}}_i \mathbf{45}^{jk}_l \overline{\mathbf{45}}_{jk}^i$, $\mathbf{5}^i \overline{\mathbf{15}}_{lj} \mathbf{45}^{jk}_i \mathbf{24}^{l}_k$, $\mathbf{5}^i \overline{\mathbf{15}}_{ij} \mathbf{45}^{jk}_l \mathbf{24}^{l}_k$, $\overline{\mathbf{5}}_i \overline{\mathbf{5}}_j \mathbf{15}^{jk} \mathbf{24}_{k}^i$, $\mathbf{5}^j \overline{\mathbf{5}}_i \mathbf{45}^{ik}_l \overline{\mathbf{45}}_{jk}^l$, and $\mathbf{45}^{ij}_k \overline{\mathbf{15}}_{jl} \mathbf{45}^{lk}_m \mathbf{24}_{i}^m$. The $2/3$ electric charge LQs are mixed through the third, fifth, and ninth contraction from this list and one more contraction of the form $\epsilon_{ijlmn} \mathbf{5}^k \mathbf{15}^{io} \mathbf{45}_k^{jl} \mathbf{45}^{mn}_o$. The LQs with the $4/3$ charge do not mix at all through any of these contractions if we neglect possible VEVs of the scalar fields that transform as $(\mathbf{1},\mathbf{3},1) (\in \mathbf{15})$ and $(\mathbf{1},\mathbf{3},0) (\in \mathbf{24})$. If that is not the case the $4/3$ electric charge LQs get mixed via third and ninth operators from the first list. Moreover, one needs to include in the analysis the following seven operators: $\mathbf{45}^{ij}_k \mathbf{24}^{k}_l \overline{\mathbf{45}}_{ij}^l$, $\mathbf{45}^{ij}_k \mathbf{24}^{l}_i \overline{\mathbf{45}}_{lj}^k$, $\overline{\mathbf{5}}_i \mathbf{15}^{jk} \overline{\mathbf{15}}_{jl} \mathbf{45}^{li}_k$, $\mathbf{5}^j \overline{\mathbf{5}}_i \mathbf{15}^{ik} \overline{\mathbf{15}}_{kj}$, $\mathbf{15}^{kj} \overline{\mathbf{15}}_{ki} \mathbf{45}^{lm}_j \overline{\mathbf{45}}_{lm}^i$, $\epsilon_{ijlmn} \mathbf{5}^i \mathbf{15}^{ko} \mathbf{45}_k^{jl} \mathbf{45}^{mn}_o$, and $\mathbf{15}^{lj} \overline{\mathbf{15}}_{ki} \mathbf{45}^{km}_j \overline{\mathbf{45}}_{lm}^i$. First five (last two) contractions from the second list generate additional contributions towards the mixing of the $1/3$ ($2/3$) electric charge LQs. 

To obtain a scenario comprising $5$-, $10$-, $24$-, and $45$-dimensional scalar representations one should replace $15$-dimensional representation with $10$-dimensional one wherever possible. Note that some of the contractions that one obtains with the simple substitution yield zero due to the antisymmetric nature of $\mathbf{10}^{ij}(=-\mathbf{10}^{ji})$ in the $SU(5)$ group space. These contractions are $\mathbf{5}^i \overline{\mathbf{10}}_{ij} \mathbf{5}^j$, $\mathbf{45}^{ij}_k \overline{\mathbf{10}}_{jl} \mathbf{45}^{lk}_i$, and $\epsilon_{ijlmn} \mathbf{5}^i \mathbf{10}^{ko} \mathbf{45}_k^{jl} \mathbf{45}^{mn}_o$. Also, one needs to add two more operators --- $\mathbf{5}^i \overline{\mathbf{10}}_{jk} \mathbf{45}^{jk}_i$ and $\mathbf{5}^i \overline{\mathbf{10}}_{lm} \mathbf{45}^{lm}_j \mathbf{24}^{j}_i$ --- that are specific for the $10$-dimensional representation case.


\bibliography{references}

\begin{thebibliography}{41}
\expandafter\ifx\csname natexlab\endcsname\relax\def\natexlab#1{#1}\fi
\expandafter\ifx\csname bibnamefont\endcsname\relax
  \def\bibnamefont#1{#1}\fi
\expandafter\ifx\csname bibfnamefont\endcsname\relax
  \def\bibfnamefont#1{#1}\fi
\expandafter\ifx\csname citenamefont\endcsname\relax
  \def\citenamefont#1{#1}\fi
\expandafter\ifx\csname url\endcsname\relax
  \def\url#1{\texttt{#1}}\fi
\expandafter\ifx\csname urlprefix\endcsname\relax\def\urlprefix{URL }\fi
\providecommand{\bibinfo}[2]{#2}
\providecommand{\eprint}[2][]{\url{#2}}

\bibitem[{\citenamefont{Chua et~al.}(2000)\citenamefont{Chua, He, and
  Hwang}}]{Chua:1999si}
\bibinfo{author}{\bibfnamefont{C.-K.} \bibnamefont{Chua}},
  \bibinfo{author}{\bibfnamefont{X.-G.} \bibnamefont{He}}, \bibnamefont{and}
  \bibinfo{author}{\bibfnamefont{W.-Y.~P.} \bibnamefont{Hwang}},
  \bibinfo{journal}{Phys. Lett.} \textbf{\bibinfo{volume}{B479}},
  \bibinfo{pages}{224} (\bibinfo{year}{2000}), \eprint{hep-ph/9905340}.

\bibitem[{\citenamefont{Mahanta}(2000)}]{Mahanta:1999xd}
\bibinfo{author}{\bibfnamefont{U.}~\bibnamefont{Mahanta}},
  \bibinfo{journal}{Phys. Rev.} \textbf{\bibinfo{volume}{D62}},
  \bibinfo{pages}{073009} (\bibinfo{year}{2000}), \eprint{hep-ph/9909518}.

\bibitem[{\citenamefont{Pati and Salam}(1973)}]{Pati:1973uk}
\bibinfo{author}{\bibfnamefont{J.~C.} \bibnamefont{Pati}} \bibnamefont{and}
  \bibinfo{author}{\bibfnamefont{A.}~\bibnamefont{Salam}},
  \bibinfo{journal}{Phys. Rev.} \textbf{\bibinfo{volume}{D8}},
  \bibinfo{pages}{1240} (\bibinfo{year}{1973}).

\bibitem[{\citenamefont{Pati and Salam}(1974)}]{Pati:1974yy}
\bibinfo{author}{\bibfnamefont{J.~C.} \bibnamefont{Pati}} \bibnamefont{and}
  \bibinfo{author}{\bibfnamefont{A.}~\bibnamefont{Salam}},
  \bibinfo{journal}{Phys. Rev.} \textbf{\bibinfo{volume}{D10}},
  \bibinfo{pages}{275} (\bibinfo{year}{1974}), \bibinfo{note}{[Erratum: Phys.
  Rev.D11,703(1975)]}.

\bibitem[{\citenamefont{Fritzsch and Minkowski}(1975)}]{Fritzsch:1974nn}
\bibinfo{author}{\bibfnamefont{H.}~\bibnamefont{Fritzsch}} \bibnamefont{and}
  \bibinfo{author}{\bibfnamefont{P.}~\bibnamefont{Minkowski}},
  \bibinfo{journal}{Annals Phys.} \textbf{\bibinfo{volume}{93}},
  \bibinfo{pages}{193} (\bibinfo{year}{1975}).

\bibitem[{\citenamefont{Georgi and Glashow}(1974)}]{Georgi:1974sy}
\bibinfo{author}{\bibfnamefont{H.}~\bibnamefont{Georgi}} \bibnamefont{and}
  \bibinfo{author}{\bibfnamefont{S.~L.} \bibnamefont{Glashow}},
  \bibinfo{journal}{Phys. Rev. Lett.} \textbf{\bibinfo{volume}{32}},
  \bibinfo{pages}{438} (\bibinfo{year}{1974}).

\bibitem[{\citenamefont{Davidson et~al.}(1994)\citenamefont{Davidson, Bailey,
  and Campbell}}]{Davidson:1993qk}
\bibinfo{author}{\bibfnamefont{S.}~\bibnamefont{Davidson}},
  \bibinfo{author}{\bibfnamefont{D.~C.} \bibnamefont{Bailey}},
  \bibnamefont{and} \bibinfo{author}{\bibfnamefont{B.~A.}
  \bibnamefont{Campbell}}, \bibinfo{journal}{Z. Phys.}
  \textbf{\bibinfo{volume}{C61}}, \bibinfo{pages}{613} (\bibinfo{year}{1994}),
  \eprint{hep-ph/9309310}.

\bibitem[{\citenamefont{Hewett and Rizzo}(1997)}]{Hewett:1997ce}
\bibinfo{author}{\bibfnamefont{J.~L.} \bibnamefont{Hewett}} \bibnamefont{and}
  \bibinfo{author}{\bibfnamefont{T.~G.} \bibnamefont{Rizzo}},
  \bibinfo{journal}{Phys. Rev.} \textbf{\bibinfo{volume}{D56}},
  \bibinfo{pages}{5709} (\bibinfo{year}{1997}), \eprint{hep-ph/9703337}.

\bibitem[{\citenamefont{Nath and Fileviez~Perez}(2007)}]{Nath:2006ut}
\bibinfo{author}{\bibfnamefont{P.}~\bibnamefont{Nath}} \bibnamefont{and}
  \bibinfo{author}{\bibfnamefont{P.}~\bibnamefont{Fileviez~Perez}},
  \bibinfo{journal}{Phys. Rept.} \textbf{\bibinfo{volume}{441}},
  \bibinfo{pages}{191} (\bibinfo{year}{2007}), \eprint{hep-ph/0601023}.

\bibitem[{\citenamefont{Dorsner et~al.}(2016)\citenamefont{Dorsner, Fajfer,
  Greljo, Kamenik, and Kosnik}}]{Dorsner:2016wpm}
\bibinfo{author}{\bibfnamefont{I.}~\bibnamefont{Dorsner}},
  \bibinfo{author}{\bibfnamefont{S.}~\bibnamefont{Fajfer}},
  \bibinfo{author}{\bibfnamefont{A.}~\bibnamefont{Greljo}},
  \bibinfo{author}{\bibfnamefont{J.~F.} \bibnamefont{Kamenik}},
  \bibnamefont{and} \bibinfo{author}{\bibfnamefont{N.}~\bibnamefont{Kosnik}},
  \bibinfo{journal}{Phys. Rept.} \textbf{\bibinfo{volume}{641}},
  \bibinfo{pages}{1} (\bibinfo{year}{2016}), \eprint{1603.04993}.

\bibitem[{\citenamefont{Shanker}(1982{\natexlab{a}})}]{Shanker:1981mj}
\bibinfo{author}{\bibfnamefont{O.~U.} \bibnamefont{Shanker}},
  \bibinfo{journal}{Nucl. Phys.} \textbf{\bibinfo{volume}{B206}},
  \bibinfo{pages}{253} (\bibinfo{year}{1982}{\natexlab{a}}).

\bibitem[{\citenamefont{Shanker}(1982{\natexlab{b}})}]{Shanker:1982nd}
\bibinfo{author}{\bibfnamefont{O.~U.} \bibnamefont{Shanker}},
  \bibinfo{journal}{Nucl. Phys.} \textbf{\bibinfo{volume}{B204}},
  \bibinfo{pages}{375} (\bibinfo{year}{1982}{\natexlab{b}}).

\bibitem[{\citenamefont{Buchmuller and Wyler}(1986)}]{Buchmuller:1986iq}
\bibinfo{author}{\bibfnamefont{W.}~\bibnamefont{Buchmuller}} \bibnamefont{and}
  \bibinfo{author}{\bibfnamefont{D.}~\bibnamefont{Wyler}},
  \bibinfo{journal}{Phys. Lett.} \textbf{\bibinfo{volume}{B177}},
  \bibinfo{pages}{377} (\bibinfo{year}{1986}).

\bibitem[{\citenamefont{Buchmuller et~al.}(1987)\citenamefont{Buchmuller,
  Ruckl, and Wyler}}]{Buchmuller:1986zs}
\bibinfo{author}{\bibfnamefont{W.}~\bibnamefont{Buchmuller}},
  \bibinfo{author}{\bibfnamefont{R.}~\bibnamefont{Ruckl}}, \bibnamefont{and}
  \bibinfo{author}{\bibfnamefont{D.}~\bibnamefont{Wyler}},
  \bibinfo{journal}{Phys. Lett.} \textbf{\bibinfo{volume}{B191}},
  \bibinfo{pages}{442} (\bibinfo{year}{1987}), \bibinfo{note}{[Erratum: Phys.
  Lett.B448,320(1999)]}.

\bibitem[{\citenamefont{Hewett and Pakvasa}(1988)}]{Hewett:1987yg}
\bibinfo{author}{\bibfnamefont{J.~L.} \bibnamefont{Hewett}} \bibnamefont{and}
  \bibinfo{author}{\bibfnamefont{S.}~\bibnamefont{Pakvasa}},
  \bibinfo{journal}{Phys. Rev.} \textbf{\bibinfo{volume}{D37}},
  \bibinfo{pages}{3165} (\bibinfo{year}{1988}).

\bibitem[{\citenamefont{Leurer}(1994{\natexlab{a}})}]{Leurer:1993em}
\bibinfo{author}{\bibfnamefont{M.}~\bibnamefont{Leurer}},
  \bibinfo{journal}{Phys. Rev.} \textbf{\bibinfo{volume}{D49}},
  \bibinfo{pages}{333} (\bibinfo{year}{1994}{\natexlab{a}}),
  \eprint{hep-ph/9309266}.

\bibitem[{\citenamefont{Leurer}(1994{\natexlab{b}})}]{Leurer:1993qx}
\bibinfo{author}{\bibfnamefont{M.}~\bibnamefont{Leurer}},
  \bibinfo{journal}{Phys. Rev.} \textbf{\bibinfo{volume}{D50}},
  \bibinfo{pages}{536} (\bibinfo{year}{1994}{\natexlab{b}}),
  \eprint{hep-ph/9312341}.

\bibitem[{\citenamefont{Aristizabal~Sierra
  et~al.}(2008)\citenamefont{Aristizabal~Sierra, Hirsch, and
  Kovalenko}}]{AristizabalSierra:2007nf}
\bibinfo{author}{\bibfnamefont{D.}~\bibnamefont{Aristizabal~Sierra}},
  \bibinfo{author}{\bibfnamefont{M.}~\bibnamefont{Hirsch}}, \bibnamefont{and}
  \bibinfo{author}{\bibfnamefont{S.~G.} \bibnamefont{Kovalenko}},
  \bibinfo{journal}{Phys. Rev.} \textbf{\bibinfo{volume}{D77}},
  \bibinfo{pages}{055011} (\bibinfo{year}{2008}), \eprint{0710.5699}.

\bibitem[{\citenamefont{Helo et~al.}(2015)\citenamefont{Helo, Hirsch, Ota, and
  Pereira~dos Santos}}]{Helo:2015fba}
\bibinfo{author}{\bibfnamefont{J.~C.} \bibnamefont{Helo}},
  \bibinfo{author}{\bibfnamefont{M.}~\bibnamefont{Hirsch}},
  \bibinfo{author}{\bibfnamefont{T.}~\bibnamefont{Ota}}, \bibnamefont{and}
  \bibinfo{author}{\bibfnamefont{F.~A.} \bibnamefont{Pereira~dos Santos}},
  \bibinfo{journal}{JHEP} \textbf{\bibinfo{volume}{05}}, \bibinfo{pages}{092}
  (\bibinfo{year}{2015}), \eprint{1502.05188}.

\bibitem[{\citenamefont{Pas and Schumacher}(2015)}]{Pas:2015hca}
\bibinfo{author}{\bibfnamefont{H.}~\bibnamefont{Pas}} \bibnamefont{and}
  \bibinfo{author}{\bibfnamefont{E.}~\bibnamefont{Schumacher}},
  \bibinfo{journal}{Phys. Rev.} \textbf{\bibinfo{volume}{D92}},
  \bibinfo{pages}{114025} (\bibinfo{year}{2015}), \eprint{1510.08757}.

\bibitem[{\citenamefont{Hagedorn et~al.}(2016)\citenamefont{Hagedorn, Ohlsson,
  Riad, and Schmidt}}]{Hagedorn:2016dze}
\bibinfo{author}{\bibfnamefont{C.}~\bibnamefont{Hagedorn}},
  \bibinfo{author}{\bibfnamefont{T.}~\bibnamefont{Ohlsson}},
  \bibinfo{author}{\bibfnamefont{S.}~\bibnamefont{Riad}}, \bibnamefont{and}
  \bibinfo{author}{\bibfnamefont{M.~A.} \bibnamefont{Schmidt}},
  \bibinfo{journal}{JHEP} \textbf{\bibinfo{volume}{09}}, \bibinfo{pages}{111}
  (\bibinfo{year}{2016}), \eprint{1605.03986}.

\bibitem[{\citenamefont{Cheung et~al.}(2016)\citenamefont{Cheung, Nomura, and
  Okada}}]{Cheung:2016fjo}
\bibinfo{author}{\bibfnamefont{K.}~\bibnamefont{Cheung}},
  \bibinfo{author}{\bibfnamefont{T.}~\bibnamefont{Nomura}}, \bibnamefont{and}
  \bibinfo{author}{\bibfnamefont{H.}~\bibnamefont{Okada}},
  \bibinfo{journal}{Phys. Rev.} \textbf{\bibinfo{volume}{D94}},
  \bibinfo{pages}{115024} (\bibinfo{year}{2016}), \eprint{1610.02322}.

\bibitem[{\citenamefont{Georgi and Jarlskog}(1979)}]{Georgi:1979df}
\bibinfo{author}{\bibfnamefont{H.}~\bibnamefont{Georgi}} \bibnamefont{and}
  \bibinfo{author}{\bibfnamefont{C.}~\bibnamefont{Jarlskog}},
  \bibinfo{journal}{Phys. Lett.} \textbf{\bibinfo{volume}{B86}},
  \bibinfo{pages}{297} (\bibinfo{year}{1979}).

\bibitem[{\citenamefont{Dorsner et~al.}(2011)\citenamefont{Dorsner, Drobnak,
  Fajfer, Kamenik, and Kosnik}}]{Dorsner:2011ai}
\bibinfo{author}{\bibfnamefont{I.}~\bibnamefont{Dorsner}},
  \bibinfo{author}{\bibfnamefont{J.}~\bibnamefont{Drobnak}},
  \bibinfo{author}{\bibfnamefont{S.}~\bibnamefont{Fajfer}},
  \bibinfo{author}{\bibfnamefont{J.~F.} \bibnamefont{Kamenik}},
  \bibnamefont{and} \bibinfo{author}{\bibfnamefont{N.}~\bibnamefont{Kosnik}},
  \bibinfo{journal}{JHEP} \textbf{\bibinfo{volume}{11}}, \bibinfo{pages}{002}
  (\bibinfo{year}{2011}), \eprint{1107.5393}.

\bibitem[{\citenamefont{Dorsner et~al.}(2012)\citenamefont{Dorsner, Fajfer, and
  Kosnik}}]{Dorsner:2012nq}
\bibinfo{author}{\bibfnamefont{I.}~\bibnamefont{Dorsner}},
  \bibinfo{author}{\bibfnamefont{S.}~\bibnamefont{Fajfer}}, \bibnamefont{and}
  \bibinfo{author}{\bibfnamefont{N.}~\bibnamefont{Kosnik}},
  \bibinfo{journal}{Phys. Rev.} \textbf{\bibinfo{volume}{D86}},
  \bibinfo{pages}{015013} (\bibinfo{year}{2012}), \eprint{1204.0674}.

\bibitem[{\citenamefont{Miura}(2016)}]{Miura:2016krn}
\bibinfo{author}{\bibfnamefont{M.}~\bibnamefont{Miura}}
  (\bibinfo{collaboration}{Super-Kamiokande}) (\bibinfo{year}{2016}),
  \eprint{1610.03597}.

\bibitem[{\citenamefont{Hiller and Schmaltz}(2014)}]{Hiller:2014yaa}
\bibinfo{author}{\bibfnamefont{G.}~\bibnamefont{Hiller}} \bibnamefont{and}
  \bibinfo{author}{\bibfnamefont{M.}~\bibnamefont{Schmaltz}},
  \bibinfo{journal}{Phys. Rev.} \textbf{\bibinfo{volume}{D90}},
  \bibinfo{pages}{054014} (\bibinfo{year}{2014}), \eprint{1408.1627}.

\bibitem[{\citenamefont{Barbieri et~al.}(2016)\citenamefont{Barbieri, Isidori,
  Pattori, and Senia}}]{Barbieri:2015yvd}
\bibinfo{author}{\bibfnamefont{R.}~\bibnamefont{Barbieri}},
  \bibinfo{author}{\bibfnamefont{G.}~\bibnamefont{Isidori}},
  \bibinfo{author}{\bibfnamefont{A.}~\bibnamefont{Pattori}}, \bibnamefont{and}
  \bibinfo{author}{\bibfnamefont{F.}~\bibnamefont{Senia}},
  \bibinfo{journal}{Eur. Phys. J.} \textbf{\bibinfo{volume}{C76}},
  \bibinfo{pages}{67} (\bibinfo{year}{2016}), \eprint{1512.01560}.

\bibitem[{\citenamefont{Fileviez~Perez}(2004)}]{FileviezPerez:2004hn}
\bibinfo{author}{\bibfnamefont{P.}~\bibnamefont{Fileviez~Perez}},
  \bibinfo{journal}{Phys. Lett.} \textbf{\bibinfo{volume}{B595}},
  \bibinfo{pages}{476} (\bibinfo{year}{2004}), \eprint{hep-ph/0403286}.

\bibitem[{\citenamefont{Lazarides et~al.}(1981)\citenamefont{Lazarides, Shafi,
  and Wetterich}}]{Lazarides:1980nt}
\bibinfo{author}{\bibfnamefont{G.}~\bibnamefont{Lazarides}},
  \bibinfo{author}{\bibfnamefont{Q.}~\bibnamefont{Shafi}}, \bibnamefont{and}
  \bibinfo{author}{\bibfnamefont{C.}~\bibnamefont{Wetterich}},
  \bibinfo{journal}{Nucl. Phys.} \textbf{\bibinfo{volume}{B181}},
  \bibinfo{pages}{287} (\bibinfo{year}{1981}).

\bibitem[{\citenamefont{Mohapatra and Senjanovic}(1981)}]{Mohapatra:1980yp}
\bibinfo{author}{\bibfnamefont{R.~N.} \bibnamefont{Mohapatra}}
  \bibnamefont{and}
  \bibinfo{author}{\bibfnamefont{G.}~\bibnamefont{Senjanovic}},
  \bibinfo{journal}{Phys. Rev.} \textbf{\bibinfo{volume}{D23}},
  \bibinfo{pages}{165} (\bibinfo{year}{1981}).

\bibitem[{\citenamefont{Dorsner and Fileviez~Perez}(2005)}]{Dorsner:2005fq}
\bibinfo{author}{\bibfnamefont{I.}~\bibnamefont{Dorsner}} \bibnamefont{and}
  \bibinfo{author}{\bibfnamefont{P.}~\bibnamefont{Fileviez~Perez}},
  \bibinfo{journal}{Nucl. Phys.} \textbf{\bibinfo{volume}{B723}},
  \bibinfo{pages}{53} (\bibinfo{year}{2005}), \eprint{hep-ph/0504276}.

\bibitem[{\citenamefont{Dorsner et~al.}(2006)\citenamefont{Dorsner,
  Fileviez~Perez, and Gonzalez~Felipe}}]{Dorsner:2005ii}
\bibinfo{author}{\bibfnamefont{I.}~\bibnamefont{Dorsner}},
  \bibinfo{author}{\bibfnamefont{P.}~\bibnamefont{Fileviez~Perez}},
  \bibnamefont{and}
  \bibinfo{author}{\bibfnamefont{R.}~\bibnamefont{Gonzalez~Felipe}},
  \bibinfo{journal}{Nucl. Phys.} \textbf{\bibinfo{volume}{B747}},
  \bibinfo{pages}{312} (\bibinfo{year}{2006}), \eprint{hep-ph/0512068}.

\bibitem[{\citenamefont{Aad et~al.}(2016)}]{Aad:2015caa}
\bibinfo{author}{\bibfnamefont{G.}~\bibnamefont{Aad}} \bibnamefont{et~al.}
  (\bibinfo{collaboration}{ATLAS}), \bibinfo{journal}{Eur. Phys. J.}
  \textbf{\bibinfo{volume}{C76}}, \bibinfo{pages}{5} (\bibinfo{year}{2016}),
  \eprint{1508.04735}.

\bibitem[{\citenamefont{Khachatryan et~al.}(2016)}]{Khachatryan:2015vaa}
\bibinfo{author}{\bibfnamefont{V.}~\bibnamefont{Khachatryan}}
  \bibnamefont{et~al.} (\bibinfo{collaboration}{CMS}), \bibinfo{journal}{Phys.
  Rev.} \textbf{\bibinfo{volume}{D93}}, \bibinfo{pages}{032004}
  (\bibinfo{year}{2016}), \eprint{1509.03744}.

\bibitem[{\citenamefont{Patrignani et~al.}(2016)}]{Olive:2016xmw}
\bibinfo{author}{\bibfnamefont{C.}~\bibnamefont{Patrignani}}
  \bibnamefont{et~al.} (\bibinfo{collaboration}{Particle Data Group}),
  \bibinfo{journal}{Chin. Phys.} \textbf{\bibinfo{volume}{C40}},
  \bibinfo{pages}{100001} (\bibinfo{year}{2016}).

\bibitem[{\citenamefont{Dorsner and Mocioiu}(2008)}]{Dorsner:2007fy}
\bibinfo{author}{\bibfnamefont{I.}~\bibnamefont{Dorsner}} \bibnamefont{and}
  \bibinfo{author}{\bibfnamefont{I.}~\bibnamefont{Mocioiu}},
  \bibinfo{journal}{Nucl. Phys.} \textbf{\bibinfo{volume}{B796}},
  \bibinfo{pages}{123} (\bibinfo{year}{2008}), \eprint{0708.3332}.

\bibitem[{\citenamefont{Dorsner}(2012)}]{Dorsner:2012uz}
\bibinfo{author}{\bibfnamefont{I.}~\bibnamefont{Dorsner}},
  \bibinfo{journal}{Phys. Rev.} \textbf{\bibinfo{volume}{D86}},
  \bibinfo{pages}{055009} (\bibinfo{year}{2012}), \eprint{1206.5998}.

\bibitem[{\citenamefont{Capozzi et~al.}(2016)\citenamefont{Capozzi, Lisi,
  Marrone, Montanino, and Palazzo}}]{Capozzi:2016rtj}
\bibinfo{author}{\bibfnamefont{F.}~\bibnamefont{Capozzi}},
  \bibinfo{author}{\bibfnamefont{E.}~\bibnamefont{Lisi}},
  \bibinfo{author}{\bibfnamefont{A.}~\bibnamefont{Marrone}},
  \bibinfo{author}{\bibfnamefont{D.}~\bibnamefont{Montanino}},
  \bibnamefont{and} \bibinfo{author}{\bibfnamefont{A.}~\bibnamefont{Palazzo}},
  \bibinfo{journal}{Nucl. Phys.} \textbf{\bibinfo{volume}{B908}},
  \bibinfo{pages}{218} (\bibinfo{year}{2016}), \eprint{1601.07777}.

\bibitem[{\citenamefont{Fileviez~Perez and Murgui}(2016)}]{Perez:2016qbo}
\bibinfo{author}{\bibfnamefont{P.}~\bibnamefont{Fileviez~Perez}}
  \bibnamefont{and} \bibinfo{author}{\bibfnamefont{C.}~\bibnamefont{Murgui}},
  \bibinfo{journal}{Phys. Rev.} \textbf{\bibinfo{volume}{D94}},
  \bibinfo{pages}{075014} (\bibinfo{year}{2016}), \eprint{1604.03377}.

\bibitem[{\citenamefont{Zee}(1980)}]{Zee:1980ai}
\bibinfo{author}{\bibfnamefont{A.}~\bibnamefont{Zee}}, \bibinfo{journal}{Phys.
  Lett.} \textbf{\bibinfo{volume}{B93}}, \bibinfo{pages}{389}
  (\bibinfo{year}{1980}), \bibinfo{note}{[Erratum: Phys. Lett.B95,461(1980)]}.

\end{thebibliography}
\end{document}